\documentclass[manuscrip,nonacm]{acmart}
\AtBeginDocument{%
  }

\newcommand{\blue}[1]{\textcolor{blue}{#1}}

\newcommand{\ignore}[1]{}

\begin{document}

\title{Computing In Spintronic Memory:
A Thermal Perspective}

\author{Patrick Miller}
\email{mill9582@umn.edu}

\author{H{\"u}srev Cılasun}
\email{cilas001@umn.edu}

\author{Sachin S. Sapatnekar}
\email{sachin@umn.edu}

\author{Ulya R. Karpuzcu}
\email{ukarpuzc@umn.edu}

\affiliation{%
  \institution{\\University of Minnesota, Twin Cities}
  \city{Minneapolis}
  \state{Minnesota}
  \country{USA}}

\begin{abstract}
  Computing-in-Memory (CiM) is a promising paradigm to address the memory bottleneck constraining traditional systems. Most power-efficient CiM variants can directly perform Boolean operations in non-volatile memory arrays. Higher microarchitectural activity due to CiM, however, can significantly increase power density (power per area) and result in thermal hotspots. In this paper, we provide a quantitative thermal characterization for CiM. 
We demonstrate that (i) the temperature remains mostly uniform due to lateral thermal conduction; (ii) the temperature increases linearly with 
the number of memory cells participating in computation;
(iii) the temperature decreases linearly with the memory array size;
(iv) the memory technology dictates the power density, hence the thermal characteristics.
\end{abstract}




\maketitle

\section{Introduction}
\noindent Many emerging applications including 
artificial intelligence (AI)~\cite{aiMemWall} and genomics~\cite{bigDataGenomics}
are fundamentally constrained by the 
memory bandwidth.
Computing in Memory (CiM) is a
promising paradigm that 
can circumvent the memory bottleneck by moving computations
near or within memory.
CiM systems primarily differ by the underlying memory technology, as well as how and where in the memory hierarchy computations can be performed. 
Be it through
physically distinct 
computational 
elements adjacent to memory or architectures that task 
memory cells with computational functions, 
CiM systems can serve as accelerators
for numerous application domains\cite{nejatollahi2020cryptopim, cilasun2020crafft, imani2019floatpim, khalifa2021filtpim} 
which results in very high levels of microarchitectural activity. 

Energy, once supplied, is converted into waste heat that raises temperatures if it builds up.
The ability to transfer heat out of any active region is proportional to the area of the interface between the heat source and the path out. CiM, by construction, increases microarchitectural activity in memory, a typically low-activity system component. Systems that directly compute in memory, {\em in situ} using memory cells, can draw increased power in confined regions that exceed a thermal solution's capabilities, giving rise to hotspots. This can result in cells being subjected to higher maximum temperatures, $T_{max}$, than expected for standard memory. A higher $T_{max}$ can degrade reliability and thereby compromise correctness.

In this paper, we take the first step towards thermal characterization of CiM systems, considering an emerging class of CiM 
which can perform universal Boolean computation directly in dense memory arrays, where columns or rows can participate in parallel computation.
Hypothetically, due to the high microarchitectural activity in a confined area, this type of CiM may draw more power than can be practically removed. With denser arrays than static random access memory (SRAM)~\cite{liao2020bench}, and far less static power, the most power-efficient variants augment non-volatile memory (NVM) with computation capability~\cite{magic, magicMTJ, cram, resch2023pimcity,imani2019floatpim,nejatollahi2020cryptopim}. In the following we demonstrate that, {\em depending on the technology, even highly power-efficient parallel computation in NVM can have power densities impractical for standard cooling.} 

In a dense CiM environment
the memory and computation activity is crammed in an NVM array measuring a fraction of a square millimeter~\cite{resch2023pimcity}. 
Hence, the surface area available to extract heat generated by computation is also miniaturized. At the same time, 
computing in NVM typically
utilizes 
the physical 
array area unevenly, 
reducing the effective area for heat dissipation further.

While thermal modeling of conventional processors \cite{skadron2003hot} and memory has been the subject of previous work, including 
3D-stacked systems \cite{han2021model}, to the best of our knowledge {\em this is the first exploration of thermal effects of computing in NVM considering 
the impact of application mapping/computation scheduling, memory technology, and physical array dimensions.}
We will start our discussion with a background on computing in NVM in Section~\ref{sec:back}; continue with thermal characterization and modeling 
in Section~\ref{sec:therm}; provide a quantitative analysis using representative CiM benchmarks in Sections~\ref{sec:setup} and~\ref{sec:eval}; cover the related work in Section~\ref{sec:rel}, and conclude the paper with a summary of our findings in Section~\ref{sec:conc}.

\section{Background}
\label{sec:back}
\subsection{Computing In Non-Volatile Memory}
\noindent Higher computational activity in a compact form -- as induced by CiM operations in NVM arrays -- can pose a challenge for thermal solutions in charge of preventing excessive heating due to power draw. 
This applies to even the highly power-efficient CiM variants which fuse MRAM with computation capability~\cite{magicMTJ, hoffer2022performing, cram,resch2023pimcity}. 
Without loss of generality, in this paper we focus on computing in MRAM as a representative technology due to  
(1) the performance being closest to 
SRAM and dynamic random access memory (DRAM) among NVM alternatives; (2) the density advantage over SRAM~\cite{liao2020bench}; and the higher endurance compared to other NVM technologies~\cite{zabihi2018memory}.

\begin{figure}[tbp]
\centerline{\includegraphics[scale=0.6]{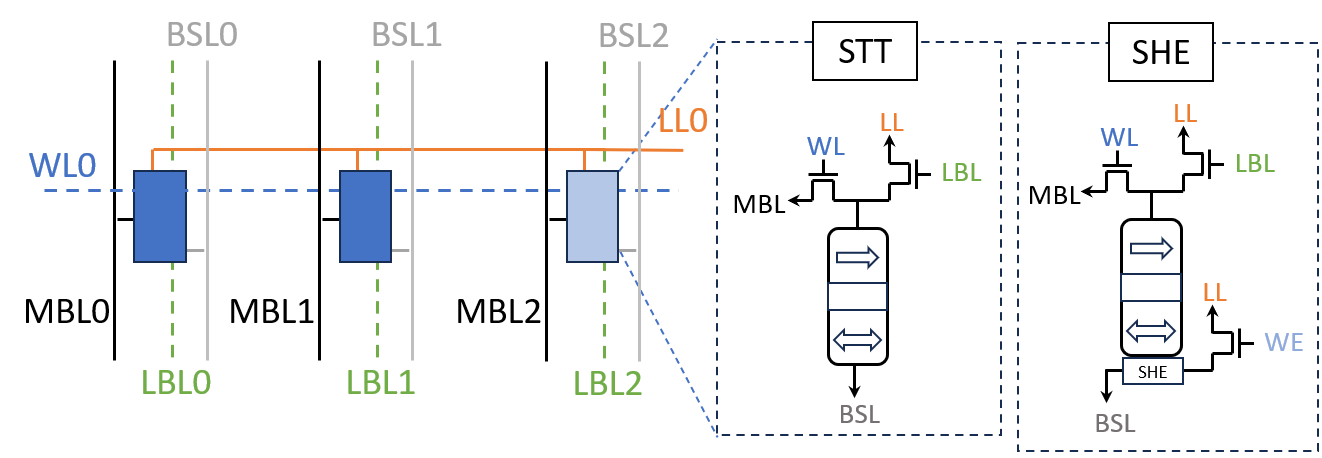}}
\caption{
Example row and cell layout with control lines to orchestrate memory and logic operations: Memory Bit Line (MBL), Bit Select Line (BSL),  
Logic Bit Line (LBL),
Logic Line (LL), and 
Word Lines (WL). 
}
\label{fig:acm_array}
\vspace{-.2cm}
\end{figure}

\begin{figure}[tbp]
\centerline{\includegraphics[scale=0.6]{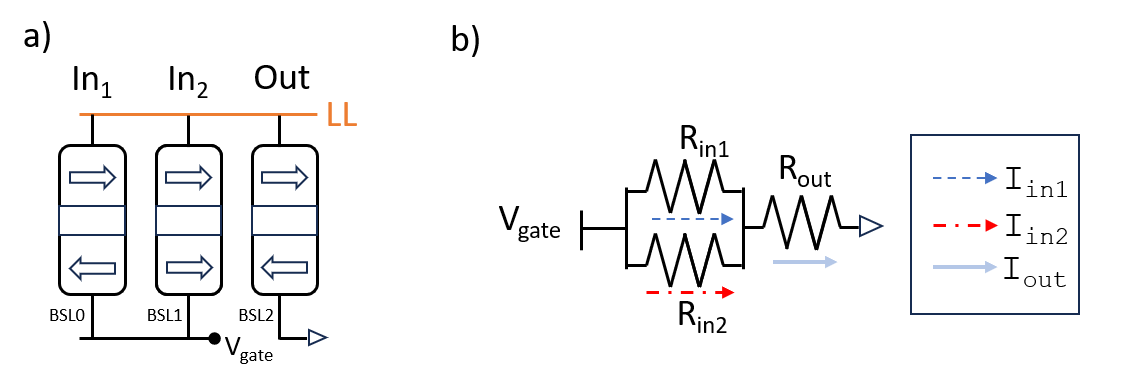}}
\caption{
(a) Connecting cells with LL for gate operation. (b) Electrical equivalent of a logic gate.
}
\label{fig:acmcircuit}
\vspace{-.2cm}
\end{figure}

Fig.\ref{fig:acm_array} and \ref{fig:acmcircuit} provide an overview for computing in MRAM. 
The CiM array behaves like a standard MRAM array if no computation takes place.
As shown in Fig.\ref{fig:acm_array}, each MRAM cell incorporates a magnetic tunnel junction (MTJ) as the storage element.  
The MTJ assumes a low resistance $R_P$ if the magnetic orientation in its fixed and and free layers are parallel; and a high resistance $R_{AP}$,  if not parallel. These two distinguishable resistance levels encode logic values zero and one, respectively. 

MTJs come in two main flavors:
Spin Transfer Torque (STT), and Spin Hall Effect (SHE) / Spin Orbit Torque (SOT). 
As a memory cell, an
STT MTJ's state is modified by passing a current exceeding a threshold $I_{crit(ical)}$ that induces the free layer orientation to change based on the direction of current flow~\cite{zabihi2018memory}.
A SHE MTJ augments the MTJ stack with a SHE channel that lowers $I_{crit}$, which in turn improves the power-efficiency~\cite{zabihi2019spinhall}. For both types, the operation semantics for CiM are similar. To ease illustration, we will continue our discussion with STT-based in-MRAM computing. 

Access transistors connect the MTJ to shared bit lines when control signals are asserted, as depicted in Fig.\ref{fig:acm_array}. 
 Memory operations assert the WL (Word Line) of a row to connect its MTJs to their MBL (Memory Bit Line). Reads and writes apply a controlled voltage difference between the MBL and BSL (Bit Select Line) in each column, inducing a current through the MTJs. Reads use a low voltage that results in a lower current than $I_{crit}$, and sense its magnitude to determine the logic state. Writes use a larger voltage to produce a current above $I_{crit}$ to enforce a state change.

\begin{table}[t]
\begin{center}
\resizebox{0.5\linewidth}{!}{
\begin{tabular}{|l|l|c|ll|}
\hline
  In$_1$ ($R_{in1}$)  & In$_2$ ($R_{in2}$) & Out ($R_{out}$) & $I_{out} = I_{1} + I_{2}$ &\\
  \hline
  0 ($R_{P}$) & 0 ($R_{P}$) & 0 $\rightarrow$ 1 & $I_{00}$ & $> I_{crit}$ \\
 \hline
 0 ($R_{P}$) & 1 ($R_{AP}$) & 0 $\rightarrow$ 1 & $I_{01}$ & $> I_{crit}$ \\
  \hline
  1 ($R_{AP}$) & 0 ($R_{P}$) & 0 $\rightarrow$ 1 & $I_{10}= I_{01}$ & $ > I_{crit}$ \\
  \hline
  1 ($R_{AP}$) & 1 ($R_{AP}$) & 0 & $I_{11}$ & $< I_{crit}$ \\
\hline
\end{tabular}
}
\vspace{.1cm}
\caption{A universal (2-input nand) gate implemented in mram. The output is preset to logic 0.} 
\label{tbl:and}
\vspace{-.9cm}
\end{center}
\end{table}

\begin{figure}[tbp]
\includegraphics[scale=0.5]{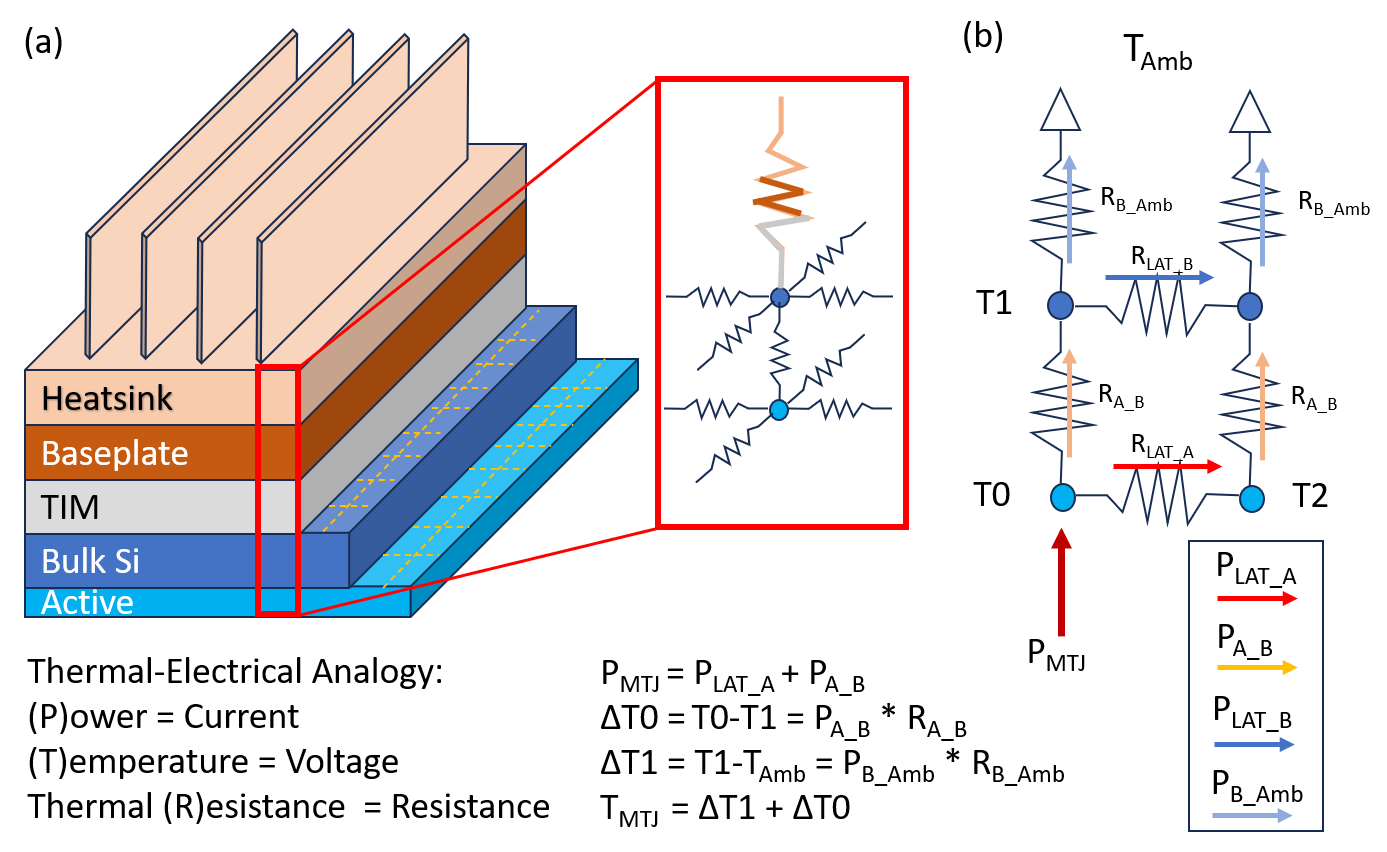}
\caption{(a) CiM modeled as a thermal resistive network. {\em Active} corresponds to a microarchitecturally active layer
(where, e.g.,  CiM takes place). 
 A layer of thermal interface
material (TIM) joins the chip with the metal baseplate
of a heatsink with fins giving a convective
interface to the environment. (b) Thermal resistance diagram of a 2D slice, with $R_{LAT\_A}$, $R_{A\_B}$, $R_{LAT\_B}$, $R_{B\_Amb}$ capturing lateral resistance in the active layer, vertical resistance between active and bulk layers, lateral resistance in the bulk layer, and the bulk layer to ambient resistance, respectively.
}
\label{fig:trn}
\vspace{-0.5cm}
\end{figure}

Numerous gates can be implemented in MRAM using this principle, including universal gates, and any cell can serve as a gate input or output. To perform a gate operation, the output cell is first preset to a gate-specific logical state. Next, a gate-specific voltage difference of $V_{gate}$ is applied between the shared BSLs of the inputs and the BSL of the output.
Finally, all inputs and output are connected to LL (Logic Line) by asserting their LBL (Logic Bit Line), forming a resistive network as demonstrated in Fig.\ref{fig:acmcircuit}(a),(b) for a representative two-input gate. 
The resistance level -- hence logic state -- of the inputs modulates the total current through the output cell. If the current exceeds $I_{crit}$, the output MTJ changes state. The choice of preset value and $V_{gate}$ constrain changes to match the truth table of the respective Boolean gate, as depicted for a 2-input NAND in Table~\ref{tbl:and}. 
SHE write and logic operations use the same overall concept, except that they have a separate access transistor controlled by a Write Enable (WE) signal that is asserted instead of LBL for the output cell, routing current to the SHE channel.
 
In this array, each row can perform one logic gate at a time, but all rows can perform the same logic gate (on different data, by construction) in parallel as the control lines span all columns in the array. This principle applies if the direction of gate formation is transposed, enabling column-parallel in-MRAM computing~\cite{resch2020mouse}. Computing in both directions (row-parallel and column-parallel) interchangeably by introducing an orthogonal set of LLs is also possible~\cite{resch2023pimcity}, and we rely on this more flexible design option in our study.  
In addition to finer grain row- or column-parallelism, coarser grain array-parallelism also applies, where
multiple MRAM arrays work concurrently. 

\subsection{Application Mapping}
\label{sec:backApp}
\noindent In-MRAM computing is Turing-complete, as any algorithm can be synthesized using 
universal Boolean gates in-MRAM. Examples include but are not limited to Spiking~\cite{cilasun2021spiking} or Binary Neural Networks~\cite{resch2020mouse}. 
The key 
is 
scheduling Boolean gates in-MRAM in time and space. The thermal 
profile is a strong function of the resulting computational activity.

As a representative example, we use Hopfield Neural Networks for end-to-end thermal benchmarking, which constitute  
fully connected networks of N neurons 
specified
by a binary 
state vector  \textbf{V} of length N, an integer bias vector \textbf{b} of the same length, and an N$\times$N symmetric integer weight matrix \textbf{W}~\cite{hopfield1982neural, davey2004high}. 
Calculating the next state of a neuron $i$
entails the dot product of \textbf{V} and 
the column $i$ in \textbf{W} and updating the $i^{th}$ entry in \textbf{V} if the dot product result exceeds a prespecified threshold. 
Each such neuron update demarcates an {\em iteration}.
The Hopfield is a recurrent network 
where {\bf V} is updated one bit per iteration until convergence ({\bf V} stabilizes).
Accordingly, we can break down 
each iteration into three steps: 
Vector-matrix multiplication, thresholding, and \textbf{V} update. 
For an in-depth discussion of implementation details we refer the reader to~\cite{resch2023pimcity} for matrix-vector multiplication; and to~\cite{cilasun2021spiking} for thresholding. 
Similar to \cite{cilasun2020crafft}, we use a write- and area-reducing optimization by coalescing much of the activity associated with vector updates in the third step, which we refer to as {\em bulk-set}. 
The first step 
dominates the overall runtime and power draw as the problem size increases.

\section{Computing in NVM: A Thermal Perspective}
\label{sec:therm}
\subsection{Microscopic View}
\label{sec:gtp}
\noindent As a result of computational activity, modern highly-integrated systems can generate significant amounts of heat in physically compact chips that must be removed before internal temperatures rise to unacceptable levels. 
To this end, 
a path for heat conduction is formed from computationally active regions
through layers of the chip, all the way to the 
heatsink that can transfer heat to the environment, as depicted
in Fig.\ref{fig:trn}. 

For a typical chip, the bulk of heat generation occurs in a thin layer of active silicon (the {\em Active} layer from Fig.\ref{fig:trn}), where the primary elements of logic and memory reside.  
As depicted in Fig.\ref{fig:trn}, a representative chip package provides a vertical path for heat conduction through several layers of the chip, 
where a layer of thermal interface material (TIM) joins the chip with the metal baseplate of the heatsink whose fins provide a convective interface with moving air to the outside environment\footnote{
Conservatively we neglect minor contributors to heat conduction such as thermally insulating oxide, package substrate, and PCB layers.}.
Some heat can travel laterally to neighboring silicon, but eventually must exit in the vertical direction through the heatsink.
To implement this, following a similar methodology to traditional thermal models~\cite{eda2008}, we divide 
the layer of active silicon encompassed by the CiM array 
into a grid of {\em thermal nodes}. 
We also define
a second layer of nodes situated at the upper side of the electrically inactive bulk silicon of the chip. 
This space subdivision 
directly maps to the MTJ cell grid of the CiM array: Each node 
resides at the center of each MTJ cell where $\Delta x$, $\Delta y$, and $\Delta z$ map to the dimensions of the cell.
%
This 
produces a system of linear equations
as such:
\begin{equation}
    G \mathbf{T}=\mathbf{P} 
    \label{eqn:cond}
\end{equation} 
$G$ is a matrix 
capturing
the thermal conductance between neighboring nodes. Thermal conductance (the inverse of thermal resistance $R$) represents the ability of a 
material to conduct heat. 
$\mathbf{T}$ is a vector where each element 
corresponds to the temperature difference between the respective node and an ambient temperature of 25$^{\circ}$C. Finally, $\mathbf{P}$ is a vector containing the profiled power draw of each node.
We use conductance values 
along each modeled direction to populate $G$; and the calculated power draw from execution traces, to populate
\textbf{P}. Then, solving Eqn.(\ref{eqn:cond}) for \textbf{T} gives each node's temperature with respect to the ambient.
$A_x = \Delta y \Delta z$ is the cross sectional area in the x direction with 
$R_x = {\Delta x}/({k_tA_x})$, 
where $k_t$ represents the material's thermal conductivity.
For z and y axes, 
$A_y=\Delta x \Delta z$ and $A_z = \Delta x \Delta y$ apply.

Active layer nodes, by construction, have 
non-zero 
elements in \textbf{P}.
We assume that the second layer is separated from the active nodes by a column of silicon with the same $A_z$ and a $\Delta z$ equal to the thickness of the die. We model the TIM and baseplate layers as vertical resistances with $A_z$ and $\Delta z$ derived from each layer's thickness. We use these and a lumped resistance for the convective interface of the heatsink as the vertical resistance between each bulk silicon node and the environment.
This finite difference based modeling method matches
the highly regular structure of the memory array. 

\subsection{Macroscopic View}

\noindent On one hand, CiM can 
mitigate
thermal issues 
by significantly reducing the 
power 
consumption of data movement~\cite{aiMemWall}, which produces heat like computation does. The power consumption of data movement scales with the distance traveled, which CiM can physically reduce by orders of magnitude. However, this reduction in distance 
can complicate
heat dissipation 
because it also reduces 
the surface area available for heat to travel out of the active layers.
This paper focuses on an in-array CiM that
leads to the shortest distances for data movement by using the memory cells directly for 
logic operations, with matching reductions in surface area for dissipation.
Typical data layout optimizations during application mapping can further shrink this surface area 
by reserving small subsets of the already highly compact memory array for specific functions or algorithmic blocks. 
%
As a result, CiM can raise power density (power per area) to a level where heat cannot be removed from the system fast enough to prevent the formation of localized thermal hotspots. 
Excessive heating can degrade MTJ reliability 
significantly \cite{zhang2020degrade, beek2022sot}, potentially compromising functional correctness. 

As explained in Section~\ref{sec:back}, each Boolean gate operation in MRAM corresponds to a resistive network from {Fig.\ref{fig:acmcircuit}}b, 
with the overall power draw (due to {\em Joule heating} through the network) characterized by
\begin{equation}
    V_{gate}^2/R_{equiv}
    \label{eqn:power}
\end{equation}
$V_{gate}$
is the gate-specific voltage difference between the bit line terminals of input cells and the bit line linked to the output cell.
$R_{equiv}$ is the 
equivalent resistance 
of the resistive gate network:
\begin{equation}
    R_{equiv} = (R_{in1} || R_{in2} ||...) + R_{out}
\end{equation} 
Each $R_{in}$ captures the resistance of an input cell, and $R_{out}$, the output cell resistance at the beginning of the gate operation.
Each element of $\mathbf{P}$ from Eqn.(\ref{eqn:cond}) captures the power draw of all CiM gate operations within the respective node.
Joule heating is the primary contributor to the power draw due to the magnitude of the currents.

\section{Evaluation Setup}
\label{sec:setup}

\noindent{\bf CiM Performance Model \& Technology Parameters:}
\begin{table*}[t]
\centering
\footnotesize
\parbox{0.23\linewidth}{
 \begin{tabular}{|c|c|c|} 
 \hline
 Parameter & STT & SHE \\ [0.5ex]
 & \cite{cilasun2024ecc} & \cite{zabihi2019spinhall} \\
 \hline\hline
 $R_P$ $(k\Omega)$ & 3.15& 253.97\\ 
 $R_{AP}$ $(k\Omega)$ & 7.34 & 507.94\\
 $R_{SHE}$ $(k\Omega)$ & - & 64\\
 $I_{crit}$ $(\mu A)$ & 50& 3\\
 $t_{sw}$ $(ns)$  & 1 & 1 \\
 \hline 
 $t_{Clk}$ $(ns)$ & 3 &3 \\[1ex]
\hline
\end{tabular}
\caption{
Technology \\Parameters
\label{tab:cellparams}}
}
\parbox{0.25\linewidth}{
\centering
\footnotesize
\begin{tabular}{|c|c|c|}
    \hline
        Gate     & STT  & SHE \\
        \hline
        Write & 0.404 & 0.211 \\
        NOR2 & 0.252 & 0.687 \\
        OR2 & 0.461  & 0.687 \\
        NAND2 & 0.304 & 0.877 \\
        AND2 & 0.514 & 0.877 \\
        MAJ3 & 0.442  & 0.570 \\
        MAJ5 & 0.411 & 0.417\\
        \hline
    \end{tabular}
    \caption{$V_{gate}$ (V) \label{tab:sample_volt}}
}
\parbox{0.3\linewidth}{
\centering
\footnotesize
\begin{tabular}{|c|c|c|}
    \hline
        Cell type & STT & SHE \\
         \hline
        Active $\Delta x$ $(\mu m)$ & 0.12 & 0.12 \\
        Active $\Delta y$ $(\mu m)$ & 0.12 & 0.24 \\
        Active $\Delta z$ $(\mu m)$ & 0.12 & 0.12 \\
        Bulk Si $\Delta z$ $(\mu m)$ & 500.0 & 500.0\\
        TIM $\Delta z$ $(\mu m)$ & 100 & 100 \\
        Cu base $\Delta z$ (mm) & 5.0 & 5.0 \\
        Si $k_t$ (W/mK) & 100.0 & 100.0 \\
        TIM $k_t$ (W/mK) & 3.0 & 3.0 \\
        Cu $k_t$ (W/mK) & 400.0 & 400.0 \\
        Convective R (C/W) & 1.5 & 1.5 \\
        Single cell area $(\mu m^2)$ & 0.0144 & 0.0288\\
        \hline
    \end{tabular}
    \caption{
    Thermal Parameters 
    }
    \label{tab:therm}
}
\vspace{-.55cm}
\end{table*}
We developed a cycle-accurate CiM simulator 
to quantitatively characterize microarchitectural activity of representative computational kernels. Simulated execution traces cover thermally significant durations for the thermal solution, or at least 1ms of simulated execution~\cite{eda2008}. 
The simulated array architecture can support both row- and column-parallel computing~\cite{resch2023pimcity}.

The simulator models cell power draw per Eq. (\ref{eqn:power}). 
We extract $V_{gate}$ for gate operations from Kirchoff's laws 
to produce the necessary $I_{crit}$ through an output cell and assume a $10\%$ voltage margin for preset/write voltage.
Table \ref{tab:sample_volt} outlines $V_{gate}$ for common instructions in the benchmark applications.
We calculate power based on the current and voltage values at each node in the network. 

Table \ref{tab:cellparams} provides the technology parameters.
STT represents the near-term technology
already on the market with promising demonstrations~\cite{IEEEIRDS2023} 
SHE primarily differs from STT in higher cell resistances, lower $I_{crit}$, and the output resistance for gate operations being replaced by the SHE channel resistance $R_{SHE}$.
$t_{sw}$ captures the switching time, the time it takes to change the MTJ state. $t_{Clk}$ on the other hand, is the clock period of the modeled architecture sufficiently long to contain the preset and switching portions of a Boolean gate operation in-MRAM.

\vspace{.1cm}
\noindent{\bf Thermal Model:}
Thermal parameters from Table \ref{tab:therm} are
based on projections for upcoming silicon processes, with moderate to conservative figures for the effectiveness of heat transport. We determine MTJ cell area 
by the footprint of the access transistors (as the magnetic and insulating components are built above the transistor stack) considering a projected leading-edge process with 30nm half-pitch\cite{liao2020bench}. 
We treat STT cells as cubes measuring 120nm on a side, with SHE doubling the length of the cell due to the second access transistor. 

We model the thermal conductivity of the active layer 
after pure silicon;
and the remaining bulk silicon, 
assuming an average wafer of 0.5mm thickness~\cite{skadron2003hot}. To be conservative, the model assumes that there is no heat transfer laterally out of the array, as if the array was surrounded by other MRAM arrays mirroring its power draw.
We also evaluate a thinned die with 50$\mu$m to quantify the impact on peak temperature.
We model the TIM (after products in use) as an inexpensive silicone/aluminum oxide grease
with a 
functional thickness of 100$\mu$m~\cite{Wallossek2024paste}. We model the heat sink 
as a device with a baseplate of the thermal conductivity of copper~\cite{skadron2003hot}, and with a bulk convective resistance to the environment 
of 0.8C/W~\cite{skadron2003hot} -- which we conservatively increase to 1.5C/W to provide some error margin in the absence of a physical setup for calibration. 

For a representative thermal characterization, we consider present and projected limits for both low- and high-end commercial technologies.
Projected thermal solutions using more conventional forced-air cooling can handle 
uniform thermal densities of $84W/cm^{2}$ in warm enclosures, where  $350W/cm^{2}$ with hotspots of $2000W/cm^{2}$ represent a limit for more exotic two-phase cooling solutions~\cite{wesling2023map}. We assume 125$^\circ$C as the limit corresponding to a maximum temperature process corner \cite{kahng2019corner}.

After constructing $\mathbf{P}$ and $\mathbf{G}$ (Section~\ref{sec:gtp}), we solve the system of equations induced by Eq.\ref{eqn:cond} using the R language and its standard math library. The main practical limitation comes from $\mathbf{G}$, a square matrix where the dimensions increase
 with the number of thermal nodes in the CiM array. To constrain this growth in the medium and large array sizes, we aggregate 4 and 8 memory cells to form a thermal node, respectively. We validate that this type of coalescing has a minor impact on thermal simulation accuracy. 

\vspace{.1cm}
\noindent{\bf Benchmarks:}
We consider 
three different categories of benchmarks to simulate representative computational activity in-MRAM, using
different problem sizes and array utilization (the ratio of actively computing rows or columns to the total number physically available). We consider three representative problem sizes for each benchmark category: {\em sm}, {\em md}, and {\em lg} for small, medium, and large, respectively. Array size, problem size, and utilization are interdependent, with the minimum size for a functioning kernel and worst-case thermal scenario being sm $(256\times32)$, while md $(512\times512)$, lg $(1024\times1024)$ are more realistic array dimensions. 

The first category is designated as a {\bf thermal virus}: INV 
has only one instruction type in its instruction mix: the invert 
operation that writes the logical inverse of the input to the output cell. 
INV features the maximum power draw compared to the other categories.
%
    INVfx implements a constant stream of INV operations to characterize the worst-case.
    INVshft captures the effect of thermal leveling 
    by rotating the affected addresses across all rows in the array, producing an even power draw throughout.

The second category is VMUL (vector-matrix multiplication), a common operation well-suited to in-MRAM computing. Without loss of generality, we use a column-parallel implementation.
%
%
The third category is the 
bipolar Hopfield network, NN, a larger scale computation for end-to-end thermal benchmarking using both row- and column-level parallelism (Section~\ref{sec:backApp}).

To quantify the impact of gate (instruction) mix on thermal behavior, VMUL and NN come in different configurations:
VMULmix and NNmix variants
    use a performance-optimized full-adder with majority gates~\cite{zabihi2018memory};
    VMULnor and NNnor, 
    a full adder implemented by solely NOR gates. 
%
%
NN configurations also differ by how state vector updates are performed (Section~\ref{sec:backApp}). As opposed to NNmixnoblk and NNnornoblk, NNmixblk and NNnorblk deploy bulk-set for parallel updates. NNmixrest and NNnorrest, on the other hand, further constrain updates to serial writes
within one row.

\section{Evaluation}
\label{sec:eval}

\subsection{Thermal Characterization of Single Array Operation}
\label{sec:tcar}

\begin{figure*}
\includegraphics[scale=1,height=5.5cm]{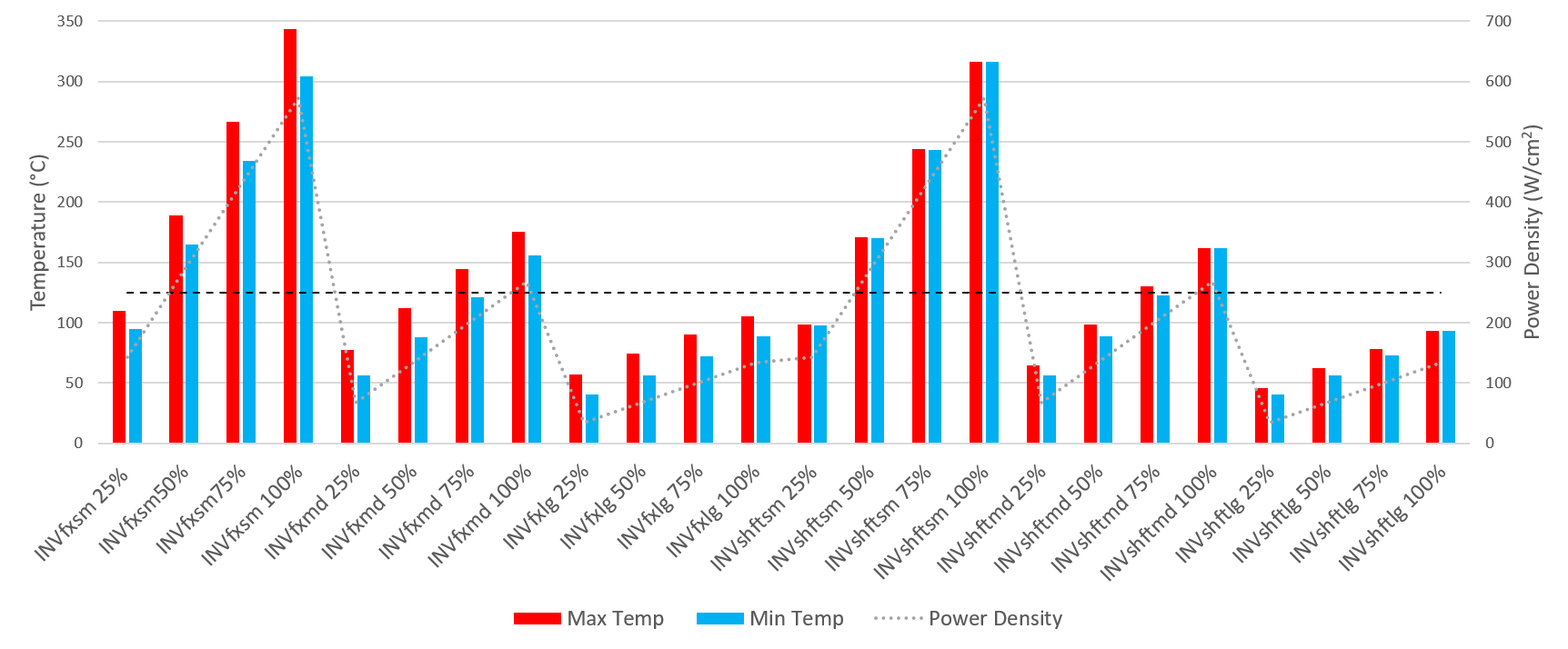}    
\caption{INV STT Temperature and Power Density. Dashed line indicates 125$^\circ$C process limit.}
\label{fig:invsttacm}
\end{figure*}

\begin{figure*}
\includegraphics[scale=1,height=5.5cm]{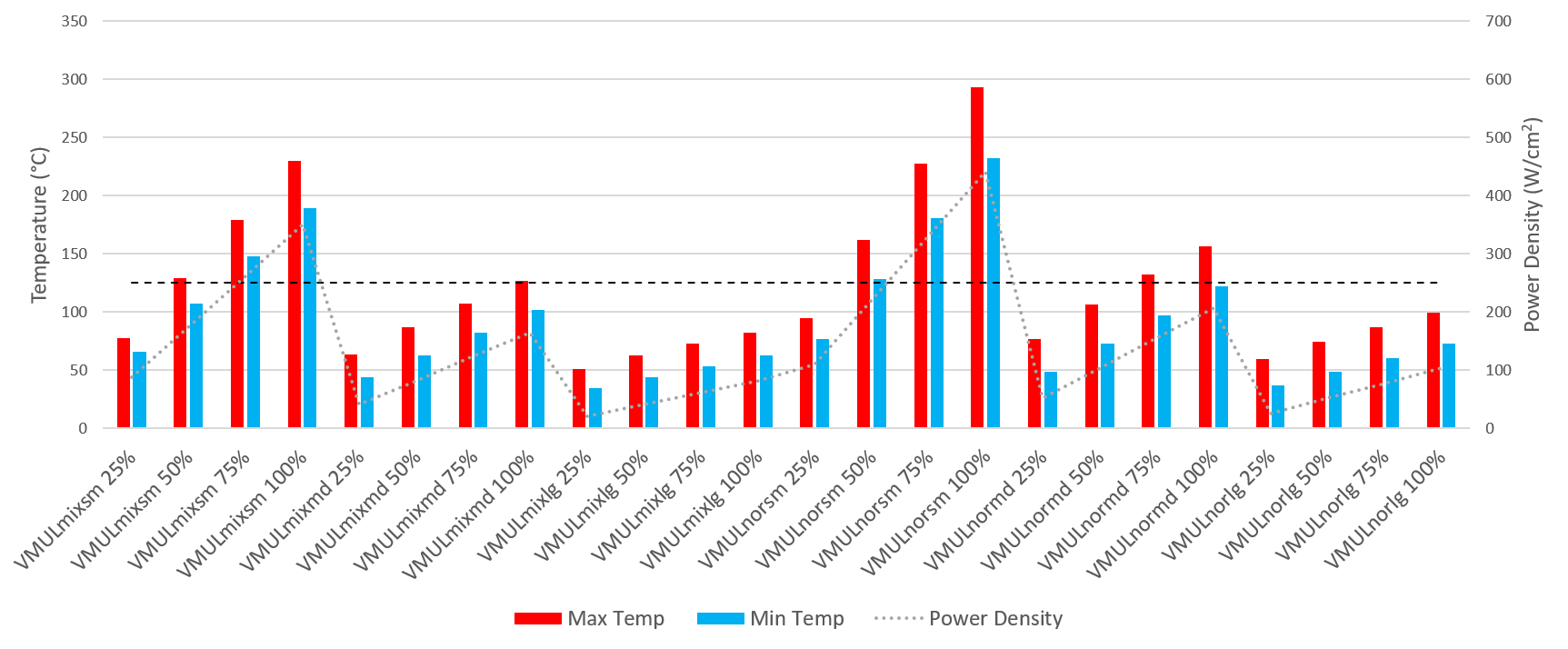}    
\caption{VMUL STT Temperature and Power Density. Dashed line indicates 125$^\circ$C process limit.}
\label{fig:vmulsttacm}
\end{figure*}

\begin{figure*}
\includegraphics[scale=1,height=5.5cm]{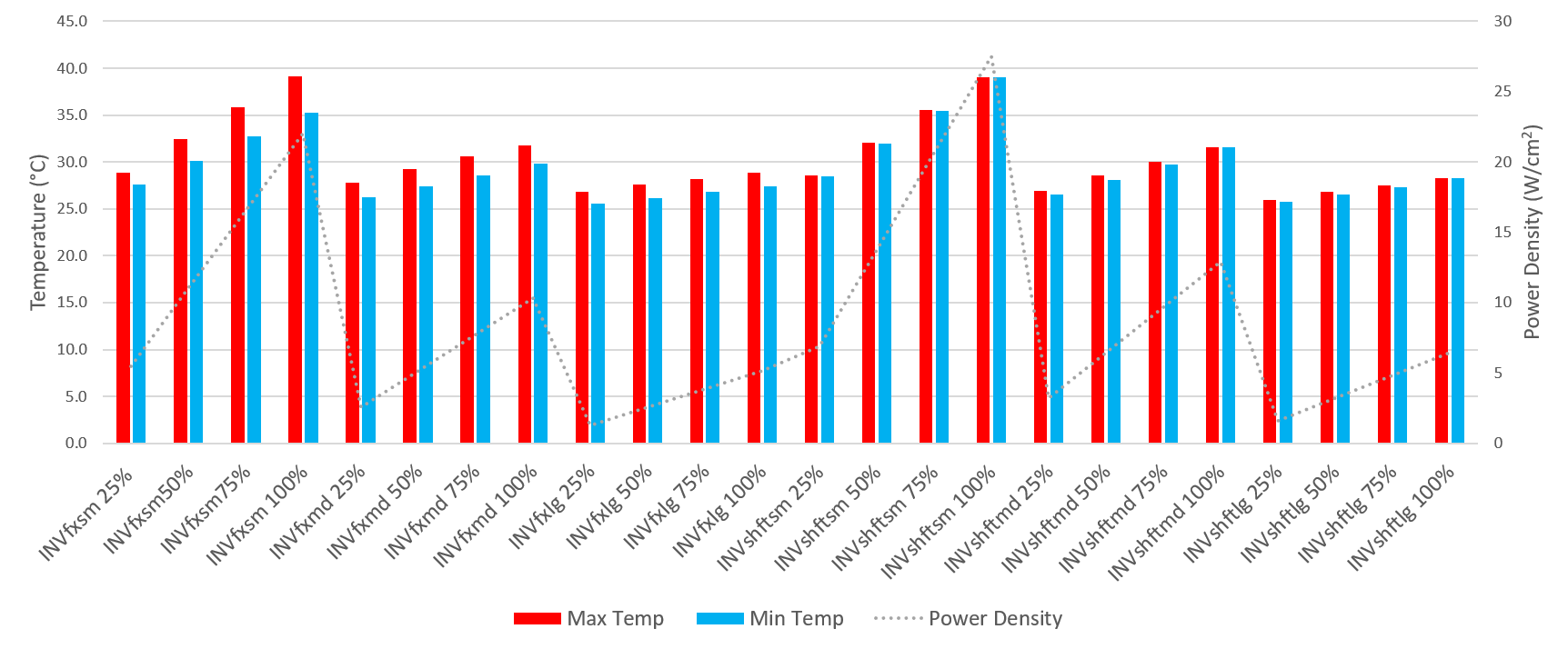}    
\caption{INV SHE Temperature and Power Density.}
\label{fig:invsheacm}
\end{figure*}

\begin{figure*}
\includegraphics[scale=1,height=5.5cm]{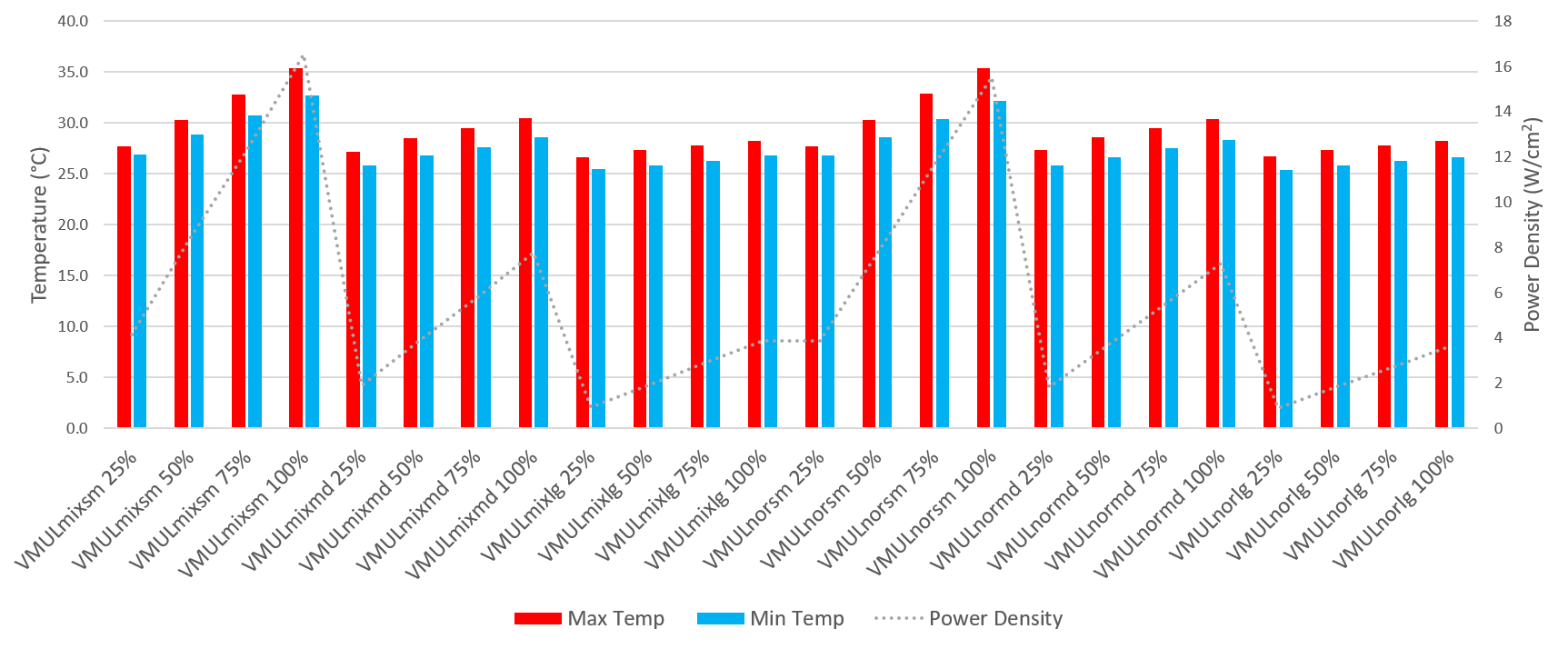}    
\caption{VMUL STT Temperature and Power Density.}
\label{fig:vmulsheacm}
\end{figure*}

\noindent We start by characterizing the maximum and minimum steady-state cell temperature across the CiM array for each benchmark configuration, as well as the power density. 
Power density is the ratio of total array power to array area, and serves as a proxy for  
the difficulty in heat dissipation relative to the limits of known cooling solutions.
{\em In this analysis, we do not assume any thermal throttling to identify the worst-case, raw thermal profile under each configuration.} Fig.\ref{fig:invsttacm} (STT) and Fig.\ref{fig:invsheacm} (SHE) show the results for INV; Fig.\ref{fig:vmulsttacm} (STT) and Fig.\ref{fig:vmulsheacm} (SHE) for VMUL; and Fig.\ref{fig:nnsttacm} (STT) and Fig.\ref{fig:nnsheacm} (SHE) for NN, respectively. Percentages on the x-axis labels capture \% utilization. Overall, we observe that:
\begin{list}{\labelitemi}{\leftmargin=1em}  
\item {\bf The temperature increases with utilization practically linearly.}  As a representative example, INVfxsm in Fig.\ref{fig:invsttacm} shows a maximum(minimum) temperature rise from 109.5(94.9)$^{\circ}$C to 343.3(304.5)$^{\circ}$C, and a power density rise from $143.1 W/cm^{2}$ to $572.5 W/cm^{2}$ as the utilization increases from 25\% to 100\%.
    The maximum(minimum) temperature rise above 25$^\circ$C goes from 85.5(69.9)$^\circ$C to 318.3(279.5)$^\circ$C, scaling
    approximately by 3.76(4.00)$\times$ where power density scales by 4.00$\times$. This very same trend applies across the board.
%
\item {\bf The temperature has an approximately 
linear inverse relationship with the array size.}
For example, INVfxmd-100\% represents an array with 34.1$\times$ larger area than INVfxsm-100\%, and has 16$\times$ more active columns, effectively 34/16=2.125$\times$ larger than INVfxmd-100\%. 
Accordingly, Fig.\ref{fig:invsttacm} reveals that the power density of the larger md configuration (INVfxmd-100\%), $268.3 W/cm^2$, is 1/2.125$\times$ of the power density of the smaller sm configuration (INVfxsm-100\%).
INVfxsm-100\% features a maximum(minimum) temperature rise of 318.3(279.5)$^\circ$C; INVfxmd-100\%, 150.8(130.5)$^\circ$C. This shows that the temperatures in the smaller sm configuration are scaled up by a factor of approximately 2.125$\times$ when compared to md. 

{There are indications that second-order effects will override this trend at dimensions beyond the current array dimensions. STT VMULnorsm overall exhibits lower peak temperatures than those INVfxsm, consistent with its lower power density, but also a larger temperature 
difference between maximum and minimum cell temperatures, which can reach 61.0$^{\circ}$C. This aligns with the above results, as VMUL places its highly localized (as well as highly active) scratch-space at the bottom of the array, leading to longer distances.}
\item {\bf With SHE, temperature increases above the ambient remain about an order of
magnitude lower than STT} due to the improved write current,
replacement of the variable output MTJ resistance with the
lower SHE channel resistance, and the doubled cell area -- as for example a direct comparison of Fig.\ref{fig:invsttacm} (STT) and Fig.\ref{fig:invsheacm} (SHE) ranges reveal for INV. The same trend applies across the board.
\end{list}

\subsection{Thermal Impact of Multi-array Operation}
\label{sec:mulArr}

\begin{figure*}
\includegraphics[scale=1,height=5.5cm]{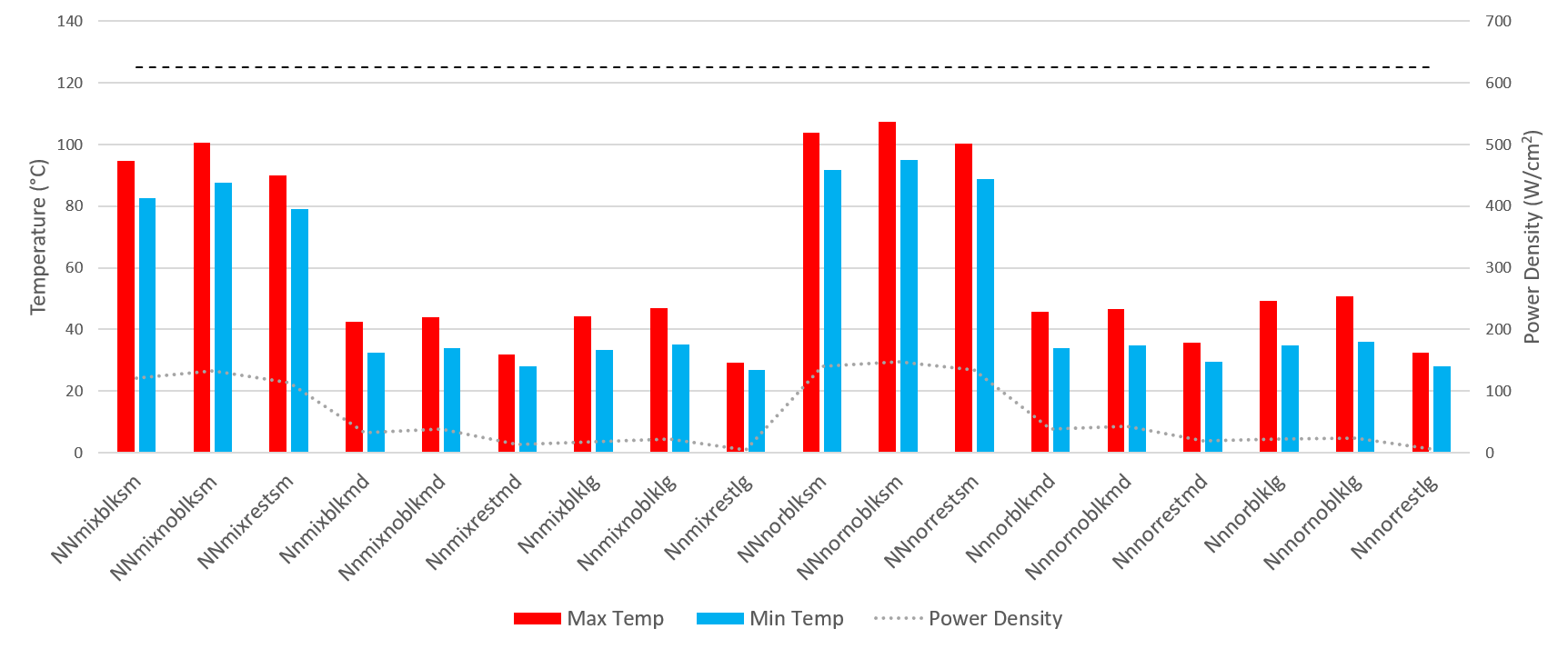}    
\caption{NN STT Temperature and Power Density. Dashed line indicates 125$^\circ$C process limit.}
\label{fig:nnsttacm}
\end{figure*}

\begin{figure*}
\includegraphics[scale=1,height=5.5cm]{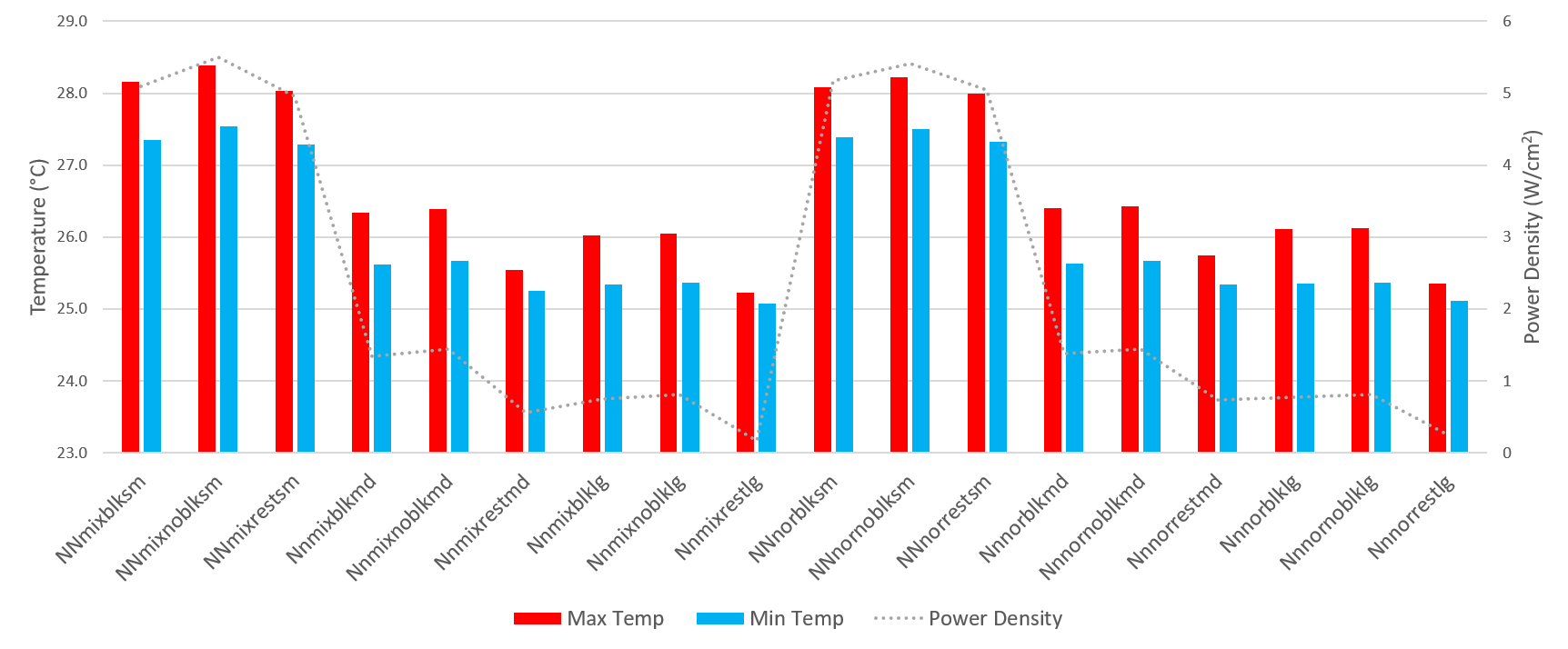}    
\caption{NN STT Temperature and Power Density.}
\label{fig:nnsheacm}
\end{figure*}

\noindent 
Fig.\ref{fig:nnsttacm} (STT) and Fig.\ref{fig:nnsheacm} (SHE) reveal for NN that 
the single array implementation results in a relatively low utilization,
leading to temperatures comparable to the lower utilization percentages of VMUL.
This is because
the single-array NN implementation  
has a highly rectangular data footprint that caps the growth of the network once it reserves the full height of a square array. 
This NN implementation is designated as a thermal benchmark to analyze the effect of heat dissipation across rows and columns. Accordingly NN needs 
four times as many rows as columns. If restricted to an array size of 512$\times$512, for example, 
we need to split the 
workload across multiple arrays, with 4 arrays mapped to the mostly column-parallel 
vector-matrix multiplication
stage, and one for the more row-parallel selection and update stage (Section~\ref{sec:backApp}). This has several benefits: The two stages can now have significantly reduced interference with the other's (especially scratch-space) activity and data; the vector-matrix multiplication stage
can utilize most of the columns (500 out of 512); and
the embarrassingly parallel column operations can concurrently run in the 4 arrays.

Our analysis reveals that, with the dominant portion of the workload split between 4 arrays,  
the overall execution time can reduce by 3.76$\times$.
The overhead mainly comes from 
the single-array row-parallel phase and 
a small number of additions to combine the parallel sums.
The upshot of increasing CiM array utilization this way is 
four times more active arrays,
which 
increases power by a factor of 3.45$\times$ (although the energy per 
iteration only increases by roughly 2\%).
In this case, the four arrays executing the column-parallel stage are effectively running a VMUL kernel on 128$\times$500 matrices, and have a power density of $133.0 W/cm^{2}$ when active. Taking into account that the arrays are inactive for approximately $20\%$ for the remaining row-parallel stage, this is comparable to 0.8$\times$ the power density of VMULmixmd-100\%, $164.3 W/cm^{2}$. 

\begin{figure}[t]
\includegraphics[scale=1,height=5.3cm]{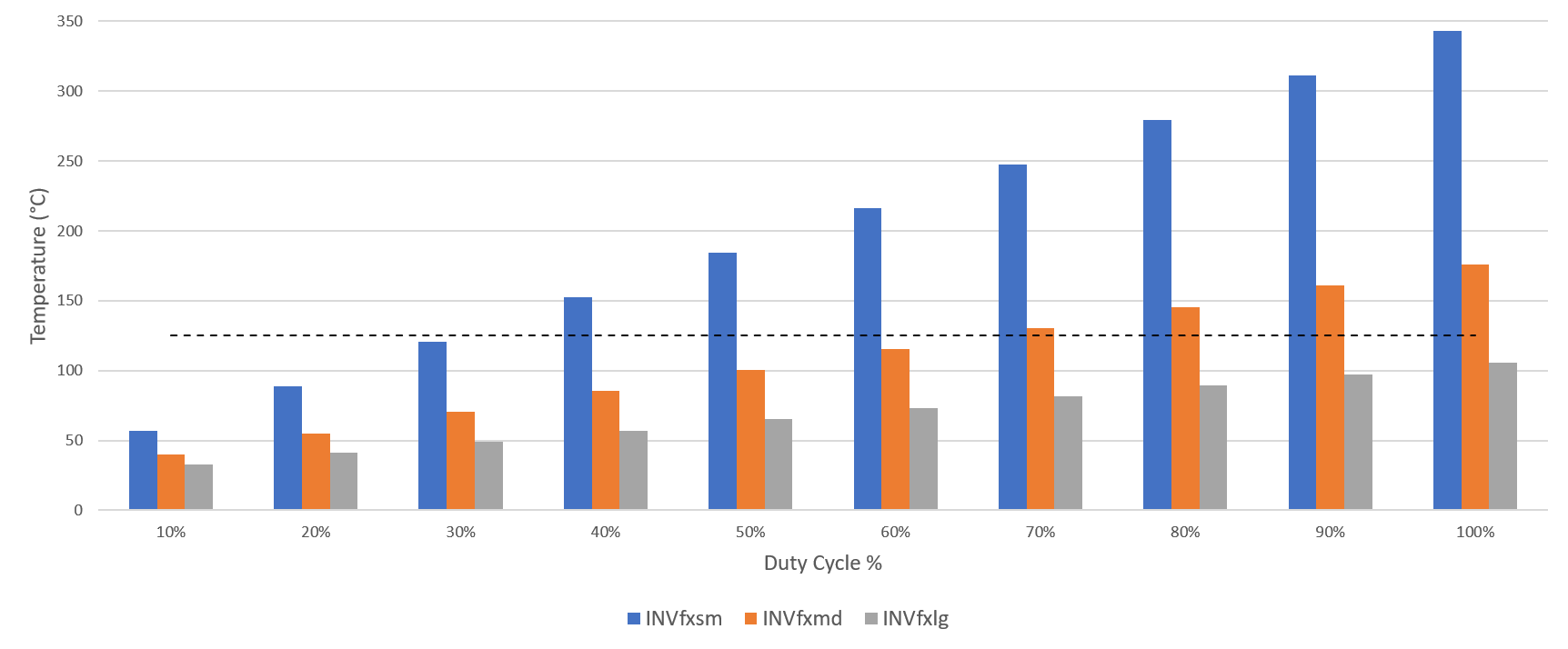}
\caption{Maximum temperature for STT INVfx-100\% under thermal throttling. Dashed line is 125$^\circ$C process limit.}
\label{fig:throttle_stt}
\vspace{-0.3cm}
\end{figure}
\subsection{Thermal Extremes}
\noindent Our analysis reveals that, for the higher-power STT technology, the benchmark configurations with the highest activity and highest thermal resistance to ambient -- predominantly INV, designated as an artificial thermal virus to identify the ultimate worst-case --
can result in unsustainable temperatures. 
In practice, thermal throttling would limit system temperature below a safe limit.
Many configurations from Fig.\ref{fig:invsttacm} and Fig.\ref{fig:vmulsttacm} exceed a typical maximum temperature corner of 125$^\circ$C for a silicon process\cite{kahng2019corner}, denoted as a dashed black line. For STT, this motivates the need for CiM-specific thermal throttling as well as a better thermal solution.

CiM power draw is driven primarily by gate operations, hence throttling by injecting idle cycles into the execution stream can linearly scale power draw at the expense of performance. Fig.\ref{fig:throttle_stt} captures this effect, where {\em duty cycle} captures the \% of non-idle cycles. 
Be it CiM-specific or not, thermal solutions feauture a rich design space.
The largest single resistance in the thermal solution is the TIM, which has many higher-cost or more complex alternatives. 
More thermally conductive materials and thinner filler layers can also help, but involve significant cost and complexity trade-offs~\cite{Wallossek2024paste}. We should also note that SHE is proving itself as a more efficient technology than STT, which is promising for adoption long-term for CiM\cite{cilasun2020crafft}.

Precautions for a thermal virus or non-ideal system conditions can be extended from existing methods for thermal throttling or enhancements to the thermal solution. The NV-CiM architecture's power draw is driven primarily by the execution of gate operations, meaning throttling by injecting a percentage of idle cycles into the execution stream can linearly scale power draw at the expense of performance. Fig. \ref{fig:throttle_stt} shows the duty cycles needed to protect the array under the power virus INV kernels.

\section{{Related Work}}
\label{sec:rel}
\noindent To the best of our knowledge, ours is the first study to characterize the thermal impact of in-memory computing by directly using memory cells for computation in situ -- as opposed to near-memory computing where computation typically happens in dedicated logic blocks or by leveraging sense amplifiers at the memory array periphery. Previous studies considered the thermal impact of in-memory computing primarily in the context of 3D stacked designs which fall under near-memory computing per our classification, and are orthogonal to our paper. 
The quality of the thermal solution has a big effect in any case, and inexpensive active heatsinks can raise 
temperature ranges significantly~\cite{han2021model}.

Our simulation and modeling methodology otherwise closely follows HotSpot~\cite{skadron2003hot}, which represents a one-of-a-kind and commonly used architecture-level thermal characterization tool. The HotSpot methodology -- of respresenting functional blocks on a silicon die as a thermal resistance network and extracting steady-state temperatures by solving differential equations derived from fundamental laws of spatial heat conduction -- equally applies in the context of in-MRAM computing
due to the uniformity of the underlying memory array.

\section{Conclusion}
\label{sec:conc}

\noindent 
Fusing the memory and compute functionality in extremely small MRAM arrays 
leads to increased microarchitectural activity coupled with significantly reduced surface area for 
heat dissipation, which can raise the operating temperature dangerously. In this paper, we provide the first study to model and characterize CiM-specific thermal effects
considering representative program execution traces and emerging CiM architectures.
We demonstrate how
power densities can exceed limits for conventional cooling, indicating the need for CiM-specific thermal throttling techniques. 
Further, we observe that temperature increases (decreases) approximately linearly with array utilization (size).
Our study also highlights the sensitivity of the thermal profile to the memory cell technology, which can result in an order of magnitude difference in power density under the same workload and architecture configuration.

\bibliographystyle{ACM-Reference-Format}
\bibliography{thermal}

@ARTICLE{aiMemWall,
author={Gholami, Amir and Yao, Zhewei and Kim, Sehoon and Hooper, Coleman and Mahoney, Michael W. and Keutzer, Kurt},
journal={ IEEE Micro },
title={{ AI and Memory Wall }},
year={2024},
volume={44},
number={03},
ISSN={1937-4143},
doi={10.1109/MM.2024.3373763},
url = {https://doi.ieeecomputersociety.org/10.1109/MM.2024.3373763},
publisher={IEEE Computer Society},
month=may}

@article{bigDataGenomics,
    doi = {10.1371/journal.pbio.1002195},
    author = {Stephens, Zachary D. AND Lee, Skylar Y. AND Faghri, Faraz AND Campbell, Roy H. AND Zhai, Chengxiang AND Efron, Miles J. AND Iyer, Ravishankar AND Schatz, Michael C. AND Sinha, Saurabh AND Robinson, Gene E.},
    journal = {PLOS Biology},
    publisher = {Public Library of Science},
    title = {Big Data: Astronomical or Genomical?},
    year = {2015},
    month = {07},
    volume = {13},
    url = {https://doi.org/10.1371/journal.pbio.1002195},
    pages = {1-11},
    number = {7}
}

@inproceedings{cilasun2020crafft,
  title={Crafft: High resolution fft accelerator in spintronic computational ram},
  author={C{\i}lasun, H{\"u}srev and Resch, Salonik and Chowdhury, Zamshed Iqbal and Olson, Erin and Zabihi, Masoud and Zhao, Zhengyang and Peterson, Thomas and Wang, Jian-Ping and Sapatnekar, Sachin S and Karpuzcu, Ulya},
  booktitle={2020 57th ACM/IEEE Design Automation Conference (DAC)},
  pages={1--6},
  year={2020},
  organization={IEEE}
}

@article{cilasun2021spiking,
  title={Spiking neural networks in spintronic computational RAM},
  author={C{\i}lasun, H{\"u}srev and Resch, Salonik and Chowdhury, Zamshed I and Olson, Erin and Zabihi, Masoud and Zhao, Zhengyang and Peterson, Thomas and Parhi, Keshab K and Wang, Jian-Ping and Sapatnekar, Sachin S and others},
  journal={ACM Transactions on Architecture and Code Optimization (TACO)},
  volume={18},
  number={4},
  pages={1--21},
  year={2021},
  publisher={ACM New York, NY, USA}
}

@inproceedings{hoffer2022performing,
  title={Performing Stateful Logic Using Spin-Orbit Torque (SOT) MRAM},
  author={Hoffer, Barak and Kvatinsky, Shahar},
  booktitle={NANO},
  year={2022},
  organization={IEEE}
}

@inproceedings{imani2019floatpim,
  title={Floatpim: In-memory acceleration of deep neural network training with high precision},
  author={Imani, Mohsen and Gupta, Saransh and Kim, Yeseong and Rosing, Tajana},
  booktitle={Proceedings of the 46th International Symposium on Computer Architecture},
  pages={802--815},
  year={2019}
}

@inproceedings{khalifa2021filtpim,
  title={FiltPIM: In-memory filter for DNA sequencing},
  author={Khalifa, Marcel and Ben-Hur, Rotem and Ronen, Ronny and Leitersdorf, Orian and Yavits, Leonid and Kvatinsky, Shahar},
  booktitle={2021 28th IEEE International Conference on Electronics, Circuits, and Systems (ICECS)},
  pages={1--4},
  year={2021},
  organization={IEEE}
}

@inproceedings{nejatollahi2020cryptopim,
  title={Cryptopim: In-memory acceleration for lattice-based cryptographic hardware},
  author={Nejatollahi, Hamid and Gupta, Saransh and Imani, Mohsen and Rosing, Tajana Simunic and Cammarota, Rosario and Dutt, Nikil},
  booktitle={2020 57th ACM/IEEE Design Automation Conference (DAC)},
  pages={1--6},
  year={2020},
  organization={IEEE}
}

@article{zabihi2018memory,
  title={In-memory processing on the spintronic CRAM: From hardware design to application mapping},
  author={Zabihi, Masoud and Chowdhury, Zamshed Iqbal and Zhao, Zhengyang and Karpuzcu, Ulya R and Wang, Jian-Ping and Sapatnekar, Sachin S},
  journal={IEEE Transactions on Computers},
  volume={68},
  number={8},
  pages={1159--1173},
  year={2018},
  publisher={IEEE}
}

@inproceedings{resch2023pimcity,
  title={PimCity: A Compute in Memory Substrate featuring both Row and Column Parallel Computing},
  author={Resch, Salonik and C{\i}lasun, H{\"u}srev and Zabihi, Masoud and Chowdhury, Zamshed and Zhao, Zhengyang and Wang, Jian-Ping and Sapatnekar, Sachin S and Karpuzcu, Ulya},
  booktitle={2023 IEEE International Conference on Rebooting Computing (ICRC)},
  pages={1--10},
  year={2023},
  organization={IEEE}
}

@inproceedings{resch2020mouse,
  title={MOUSE: Inference in non-volatile memory for energy harvesting applications},
  author={Resch, Salonik and Khatamifard, S Karen and Chowdhury, Zamshed I and Zabihi, Masoud and Zhao, Zhengyang and Cilasun, Husrev and Wang, Jian-Ping and Sapatnekar, Sachin S and Karpuzcu, Ulya R},
  booktitle={2020 53rd Annual IEEE/ACM International Symposium on Microarchitecture (MICRO)},
  pages={400--414},
  year={2020},
  organization={IEEE}
}

@article{magic,
  title={Logic design within memristive memories using memristor-aided loGIC (MAGIC)},
  author={Talati, Nishil and others},
  journal={IEEE Transactions on Nanotechnology},
  volume={15},
  number={4},
  year={2016}
}

@article{cram,
  title={Efficient in-memory processing using spintronics},
  author={Chowdhury, Zamshed and others},
  journal={IEEE CAL},
  volume={17},
  number={1},
  year={2017}
}

@inproceedings{magicMTJ,
  title={Performing memristor-aided logic (MAGIC) using STT-MRAM},
  author={Louis, Jeffry and others},
  booktitle={ICECS},
  year={2019}
}

@inproceedings{hopfield1982neural,
  title={ Neural networks and physical systems with emergent col-
lective computational abilities },
  author={Hopfield, J J },
  booktitle={ Proceedings of the National Academy of Sciences of the United States of America },
  pages={2554--2558},
  year={1982},
  month={April},
  volume= {79},
  doi = {10.1073/pnas.79.8.2554},
  organization={PNAS}
}

@article{davey2004high,
  title={High capacity associative memory models-binary and bipolar representation},
  author={Davey, Neil and Frank, Ray and Hunt, Steve and Adams, RG and Calcraft, Lee},
  journal={Procs of ASC 2004},
  year={2004}
}

@inproceedings{zabihi2019spinhall,
  title={ Using Spin-Hall MTJs to Build an Energy-Efficient
In-memory Computation Platform},
  author={Zabihi, Masoud and Zhao, Zhengyang and DC, Mahendra and Chowdhury, Zamshed and Resch, Salonik and Peterson, Thomas and Karpuzcu, Ulya R and Wang, Jian-Ping and Sapatnekar, Sachin S },
  booktitle={ 20th International Symposium on Quality Electronic Design (ISQED)},
  pages={52--57},
  year={2019},
  month={March},
  doi = { 10.1109/ISQED.2019.8697377 },
  organization={IEEE}
}

@inproceedings{cilasun2024ecc,
  title={ On Error Correction for Nonvolatile Processing-In-Memory},
  author={ Cılasun, Hüsrev and Resch, Salonik and Chowdhury, Zamshed I and Zabihi, Masoud and Lv, Yang and Zink, Brandon and Wang, Jian-Ping and Sapatnekar, Sachin S and Karpuzcu, Ulya R },
  booktitle={2024 ACM/IEEE 51st Annual International Symposium on Computer Architecture (ISCA)},
  pages={678--692},
  year={2024},
  month={June},
  doi = { 10.1109/ISCA59077.2024.00055 },
  organization={IEEE}
}

@inproceedings{skadron2003hot,
author = {Skadron, Kevin and Stan, Mircea R. and Huang, Wei and Velusamy, Sivakumar and Sankaranarayanan, Karthik and Tarjan, David},
title = {Temperature-aware microarchitecture},
year = {2003},
isbn = {0769519458},
publisher = {Association for Computing Machinery},
address = {New York, NY, USA},
url = {https://doi.org/10.1145/859618.859620},
doi = {10.1145/859618.859620},
booktitle = {Proceedings of the 30th Annual International Symposium on Computer Architecture},
pages = {2–13},
numpages = {12},
location = {San Diego, California},
series = {ISCA '03}
}

@article{liao2020bench,
  author={Liao, Yu-Ching and Pan, Chenyun and Naeemi, Azad},
  journal={IEEE Journal on Exploratory Solid-State Computational Devices and Circuits}, 
  title={Benchmarking and Optimization of Spintronic Memory Arrays}, 
  year={2020},
  volume={6},
  number={1},
  pages={9-17},
  doi={10.1109/JXCDC.2020.2999270}}

@article{zhang2020degrade,
title = {Life-time degradation of STT-MRAM by self-heating effect with TDDB model},
journal = {Solid-State Electronics},
volume = {173},
pages = {107878},
year = {2020},
issn = {0038-1101},
doi = {https://doi.org/10.1016/j.sse.2020.107878},
url = {https://www.sciencedirect.com/science/article/pii/S0038110120303452},
author = {Xue Zhang and Guangjun Zhang and Lijie Shen and Pingping Yu and Yanfeng Jiang}
}

@inproceedings{beek2022sot,
  author={Van Beek, Simon and Cai, Kaiming and Rao, Siddharth and Jayakumar, Ganesh and Couet, Sebastien and Jossart, Nico and Chasin, Adrian and Kar, Gouri Sankar},
  booktitle={2022 IEEE International Reliability Physics Symposium (IRPS)}, 
  title={MTJ degradation in SOT-MRAM by self-heating-induced diffusion}, 
  year={2022},
  volume={},
  number={},
  pages={4A.2-1-4A.2-4},
  doi={10.1109/IRPS48227.2022.9764459}}

@inproceedings{han2021model,
  author={Han, Jun-Han and West, Robert E. and Torres-Castro, Karina and Swami, Nathan and Khan, Samira and Stan, Mircea},
  booktitle={2021 International Symposium on Devices, Circuits and Systems (ISDCS)}, 
  title={Power and Thermal Modeling of In-3D-Memory Computing}, 
  year={2021},
  volume={},
  number={},
  pages={1-4},
  doi={10.1109/ISDCS52006.2021.9397913}}

@article{wesling2023map,
    author = {Wesling, Paul (Ed)},
    title = {Heterogeneous integration roadmap 2023
edition Chap 20},
    journal = {Heterogeneous Integration Roadmap},
    year = {2023}
}

@article{eda2008,
url = {http://dx.doi.org/10.1561/1000000007},
year = {2008},
volume = {2},
journal = {Foundations and Trends® in Electronic Design Automation},
title = {Thermally Aware Design},
doi = {10.1561/1000000007},
issn = {1551-3939},
number = {3},
pages = {255-370},
author = {Yong Zhan and Sanjay V. Kumar and Sachin S. Sapatnekar}
}

@techreport{IEEEIRDS2023,
  author = {{IEEE International Roadmap for Devices and Systems (IRDS)}},
  title = {2023 IEEE International Roadmap for Devices and Systems (IRDS)},
  year = {2023},
  institution = {IEEE},
  url = {$https://irds.ieee.org/images/files/pdf/2023/2023IRDS\_BC.pdf$},
}

@misc{Wallossek2024paste, 
title={Best thermal paste database and charts - paste versus paste comparison, reviews and durability for CPU and GPU}, 
url={https://www.igorslab.de/en/the-worlds-first-interactive-thermal-paste-database-real-measurement-data-material-analysis-and-objective-fact-chec}, 
journal={igor´sLAB}, 
author={Wallossek, Igor}, 
year={2024}, month={Nov}}

@INPROCEEDINGS{kahng2019corner,
  author={Kahng, Andrew B. and Mallappa, Uday and Saul, Lawrence and Tong, Shangyuan},
  booktitle={2019 Design, Automation \& Test in Europe Conference \& Exhibition (DATE)}, 
  title={"Unobserved Corner" Prediction: Reducing Timing Analysis Effort for Faster Design Convergence in Advanced-Node Design}, 
  year={2019},
  volume={},
  number={},
  pages={168-173},
  doi={10.23919/DATE.2019.8715102}}



\end{document}